\documentstyle[prb,tighten,aps,epsf]{revtex}
\begin{document}
\draft
\title{Shot noise in ferromagnetic single electron tunneling devices}
\author {B.R. Bu{\l}ka$^1$, J. Martinek$^1$, G. Micha{\l}ek$^1$ and J. Barna\'s$^2$ }
\address{$^1$Institute of Molecular Physics,
Polish Academy of Sciences,   \\
ul. Smoluchowskiego 17, 60-179 Pozna\'n, Poland\\
$^2$Department of Physics, A. Mickiewicz University, \\
     ul. Umultowska 85, 61-614 Pozna\'n, Poland}
\date{Received \hspace{5mm} }
\date{\today}
\maketitle

\begin{abstract}
 Frequency dependent current noise in ferromagnetic double junctions
with Coulomb blockade is studied theoretically in the limit of sequential
tunneling.
Two different relaxation processes are found in the
correlations between spin polarized tunneling currents;
low frequency
spin fluctuations and high frequency charge fluctuations.
Spin accumulation in strongly asymmetric junctions is shown to lead to
a negative differential resistance.
We also show that large spin noise activated in
the range of negative differential resistance gives rise to
a significant enhancement of the current noise.
\end{abstract}

\section{ Introduction}

Recent progress in nanotechnology of magnetic materials renewed interest
in spin-polarized electron tunneling in magnetic junctions.  This
interest is additionally stimulated by expected  applications of magnetic
junctions in field sensors, magnetic random access memories (MRAMs),
and other microelectronic devices.  It is well known since the work of
Julliere~\cite{jul},
that the junction resistance depends on its magnetic configuration, and
this effect, know also as the tunnel magnetoresistance (TMR), offers the
possibility of magnetic control of the tunneling current.

A special kind of tunnel junctions are the double barrier junctions
with a small central electrode. The increase in electrostatic energy
due to charging this electrode with a single electron can lead to
Coulomb blockade of electric current below a certain threshold voltage
and to Coulomb steps in I-V characteristics at higher voltages. Such
junctions, known as single electron transistors (SETs), were
extensively studied in the last decade in the limit of nonmagnetic
(metallic or semiconducting) electrodes.~\cite{SET} However, it is only very
recently when the interplay of charge and spin effects in SETs with
ferromagnetic electrodes was studied theoretically and 
experimentally.~\cite{tmr,bar,tak,sac}
It has been shown that the charging effects give rise to periodic
modulations of the TMR effect with increasing bias voltage. It has
been also predicted  that cotunneling effects in the Coulomb blockade
regime can significantly enhance TMR.
When intrinsic spin relaxation
time on the central electrode (i.e. spin relaxation due to spin-flip
scattering within the electrode) is much longer than injection time
(time between successive tunneling events), then a nonequilibrium
spin distribution of electrons is driven by the flowing current.
The corresponding spin accumulation can lead to new effects, like
negative differential resistance (NDR) or inverse TMR effect.

    A  very important parameter, in view of potential applications
of the TMR effect, is the noise to signal ratio. In the case of
nonmagnetic SETs this problem was recently extensively studied in
the relevant literature.~\cite{klu,but,jong,kor,her,oth,li,lie,li2,birk}
It has been 
shown that the noise can provide
additional information on the electronic structure, transport
properties and also on electron-electron interactions. A special issue
is the problem of suppression of the zero-frequency shot noise by
electron correlations. Negative correlations between current pulses,
induced by the Pauli principle, can lead to complete suppression of
the shot noise in quantum point contacts or to reduction of the
noise to $(1/3)S_{Poisson}$ ($S_{Poisson}=2eI$) in metallic
diffusive conductors.~\cite{klu,but,jong} Theoretical studies of current 
correlations in the presence of
Coulomb blockade in SETs by Korotkov~\cite{kor}, Hershfield et al.~\cite{her} and
others~\cite{oth} showed that the shot noise is reduced to $(1/2)S_{Poisson}$.
These predictions were also confirmed experimentally on quantum point
contacts, diffusive conductors, resonant tunneling diodes
and SETs.~\cite{li,lie,li2,birk}

    However, the situation if ferromagnetic junctions significantly
differs from that in nonmagnetic SETs and to our best knowledge there
is neither experimental nor theoretical work published up to now on
shot noise in ferromagnetic SETs. As we show in this paper, large
spin fluctuations, which occur in ferromagnetic junctions, have a
significant influence on the shot noise, particularly at low frequencies.
We also show that two different relaxation times (spin and charge) appear
in the frequency dependent current noise.  Large spin fluctuations also occur
in nonmagnetic junctions, but they do not contribute to the current noise.
Another important result following from our analysis is the presence of
an enhanced shot noise (well above  $2eI$) in the NDR range.
Similar enhancement was recently observed in
resonant tunneling diodes. The NDR is due to spin accumulation, which
in turn follows from detailed balance of electrons with opposite spins
tunneling to and off the central electrode.

    The paper is organized as follows. In Section 2 we describe
the formalism which we use for shot noise calculations in ferromagnetic
SET's with the central electrode small enough so the effects due to
charging and discrete energy levels are important.
The formalism is an extension of the formalism used for nonmagnetic
SET's by including spin degrees of freedom.
In section 3 we analyze the charge-charge and spin-spin correlations
as well as the current shot noise in ferromagnetic
junctions with a nonmagnetic central electrode.
The case when one external electrode is made of a strong ferromagnet
while the other electrodes are nonmagnetic is considered in Section 4.
 Final conclusions and
discussion are in section 5.

\section{Description of model and method of calculations}

We consider a double junction in which a nonmagnetic metallic grain
is separated from two
ferromagnetic leads (source and sink electrodes)
by tunnel barriers with the corresponding
spin dependent resistances $R_{1\uparrow}$ and $R_{1\downarrow}$ on the right
and $R_{2\uparrow}$ and  $R_{2\downarrow}$ on the left sides.
If the resistances $R_{j\sigma}$ (j=1,2) are much larger than the quantum resistance
$R_Q = h/2e^2$, electronic transport is dominated by incoherent, sequential
tunneling processes and can be described within the orthodox approach,~\cite{orth}
in which tunneling is treated in the lowest-order perturbation theory whereas
higher-order tunneling processes (cotunneling) are neglected. To describe
tunneling current and shot noise we extend previous descriptions by including
the spin degrees of freedom. This generalization is not trivial and
leads to effects which are absent in the spinless descriptions.

Electronic transport through the system is govern by the master equation
(written in the matrix form)
\begin{equation}\label{1}
\frac{d \hat{p}}{dt}=\hat{M}\hat{p}
\end{equation}
for the time evolution of the probability $p(N_{\uparrow},N_{\downarrow};t)$ to find
$N_{\uparrow}$ and $N_{\downarrow}$ excess electrons on
the grain. In the limit of long intrinsic spin relaxation on the grain
(much longer than the time between successive tunneling events)
and for spin conserving tunneling processes,
the matrix $\hat{M}$ is given by
\begin{eqnarray}\label{2}
M_{N'_{\uparrow}N'_{\downarrow};N_{\uparrow}N_{\downarrow}}=
\left\{\begin{array}{lclll}
A(N_{\uparrow},N_{\downarrow})&\mbox{for}&N'_{\uparrow}=N_{\uparrow}&
\mbox{and}&N'_{\downarrow}=N_{\downarrow}\;,\\
B^{\pm}_{\uparrow}(N_{\uparrow},N_{\downarrow})&\mbox{for}&N'_{\uparrow}=
N_{\uparrow}\pm 1&\mbox{and}&N'_{\downarrow}=N_{\downarrow}\;,\\
B^{\pm}_{\downarrow}(N_{\uparrow},N_{\downarrow})&\mbox{for}&N'_{\uparrow}=
N_{\uparrow}&\mbox{and}&N'_{\downarrow}=N_{\downarrow}\pm 1\;,\\
0&\mbox{otherwise}\;,&&&
\end{array}\right.
\end{eqnarray}
where
\begin{eqnarray}\label{2a}
A(N_{\uparrow},N_{\downarrow})=-\Gamma^+_{1\uparrow}(N_{\uparrow},N_{\downarrow})-
\Gamma^-_{1\uparrow}(N_{\uparrow},N_{\downarrow})-
\Gamma^+_{1\downarrow}(N_{\uparrow},N_{\downarrow})-
\Gamma^-_{1\downarrow}(N_{\uparrow},N_{\downarrow})\nonumber\\
-\Gamma^+_{2\uparrow}(N_{\uparrow},N_{\downarrow})-
\Gamma^-_{2\uparrow}(N_{\uparrow},N_{\downarrow})-\Gamma^+_{2\downarrow}(N_{\uparrow},N_{\downarrow})-
\Gamma^-_{2\downarrow}(N_{\uparrow},N_{\downarrow})\;,\\
B^{\pm}_{\uparrow}(N_{\uparrow},N_{\downarrow})=\Gamma^{\pm}_{1\uparrow}(N_{\uparrow}\pm1,N_{\downarrow})+
\Gamma^{\pm}_{2\uparrow}(N_{\uparrow}\pm1,N_{\downarrow})\;.\\
B^{\pm}_{\downarrow}(N_{\uparrow},N_{\downarrow}) =\Gamma^{\pm}_{1\downarrow}(N_{\uparrow},N_{\downarrow}\pm1)+
\Gamma^{\pm}_{2\downarrow}(N_{\uparrow},N_{\downarrow}\pm1)\;.
\end{eqnarray}
The tunneling rates for electrons with spin $\sigma$, tunneling to ($+$)
and off ($-$) the grain through the $j$-th junction, are given by~\cite{bee}
\begin{eqnarray}\label{3}
\Gamma^{\pm}_{j\sigma}(N_{\uparrow},N_{\downarrow})=\frac{\Delta E}{e^2R_{j\sigma}}
\sum_{i} \left[1+\exp \left(\mp\frac{E_{i\sigma}-\Delta E N_{\sigma}-E_F}{\frac{1}{2}k_BT}\right)\right]^{-1}\nonumber\\
\left[1+\exp \left(\pm\frac{E_{i\sigma}+eV_{j}(N_{\uparrow},N_{\downarrow})\mp E_c-E_F}{k_BT}\right)\right]^{-1}
\;,
\end{eqnarray}
where $E_F$ denotes the Fermi energy and
the summation runs over discrete energy levels $E_{i\sigma}$ of the grain,
which are assumed to be equally separated with the level spacing $\Delta E$.
Apart from this,
\begin{equation}\label{3a}
V_j(N_{\uparrow},N_{\downarrow})=(-1)^jV\frac{C_1C_2}{C_jC}+\frac{e N}{C}
\end{equation}
and $N= N_{\uparrow}+N_{\downarrow}$ is the total number
of excess electrons on the grain, $C_{1,2}$ are the capacitances of
the right of and left junctions,
$C$ is the total capacitance $C=C_1+C_2$, while
$e$ and $T$ stand for the electron charge and temperature, respectively.
The single-electron charging energy is defined as $E_c=e^2/2C$. When writing
Eq.(6) we assumed that
electron energy relaxation time is of the order or shorter
than the time between successive tunneling events, so electrons of a given
spin orientation are in thermal equilibrium.

The electric current in the stationary state can be
calculated from the formula
\begin{eqnarray}\label{4}
I= e\sum_{N_{\uparrow},N_{\downarrow}}[\Gamma^+_{1\uparrow}(N_{\uparrow},
N_{\downarrow})-\Gamma^-_{1\uparrow}(N_{\uparrow},N_{\downarrow}) +
\Gamma^+_{1\downarrow}(N_{\uparrow},N_{\downarrow})-
\Gamma^-_{1\downarrow}(N_{\uparrow},N_{\downarrow})]p^0(N_{\uparrow},N_{\downarrow})\nonumber \\
=e\sum_{N_{\uparrow},N_{\downarrow}}[\Gamma^-_{2\uparrow}(N_{\uparrow},N_{\downarrow})-
\Gamma^+_{2\uparrow}(N_{\uparrow},N_{\downarrow}) + \Gamma^-_{2\downarrow}(N_{\uparrow},N_{\downarrow})-
\Gamma^+_{2\downarrow}(N_{\uparrow},N_{\downarrow})]p^0(N_{\uparrow},N_{\downarrow})\;,
\end{eqnarray}
where the probability $p^0(N_{\uparrow},N_{\downarrow})$ is a solution
of the equation $\hat{0}=\hat{M}\hat{p}^0$. One can also determine the
average value of any physical quantity $X$ in the stationary state as
\begin{equation}\label{4a}
<X> = \sum_{N_{\uparrow},N_{\downarrow}}
X_{N_{\uparrow},N_{\downarrow}}p^0(N_{\uparrow},N_{\downarrow})\;,
\end{equation}
where $X_{N_{\uparrow},N_{\downarrow}}$ is the representation of $X$
in the space of states described by
the numbers $N_{\uparrow}$ and $N_{\downarrow}$
of excess electrons on the grain.

To analyze fluctuations in the system we extend the generation-recombination
approach~\cite{vliet} for multi-electron channels by generalization of the
method developed for spinless electrons in SET~\cite{kor,her,oth}.
The time correlation function of the quantities
$X$ and $Y$ is expressed as~\cite{vliet}
\begin{equation}\label{5}
<X(t)Y(0)> = \sum_{N'_{\uparrow},N'_{\downarrow};N_{\uparrow},N_{\downarrow}}
X_{N'_{\uparrow},N'_{\downarrow}}P(N'_{\uparrow},N'_{\downarrow};t|N_{\uparrow},N_{\downarrow};0)
Y_{N_{\uparrow},N_{\downarrow}}p^0(N_{\uparrow},N_{\downarrow})\;.
\end{equation}
Here, $P(N'_{\uparrow},N'_{\downarrow};t|N_{\uparrow},N_{\downarrow};0)$
is the conditional probability to find the system
in the final state with $N'_{\uparrow}$ and $N'_{\downarrow}$
excess electrons at time $t$, if
there was $N_{\uparrow}$ and  $N_{\downarrow}$ excess electrons
in the initial time t=0.
This probability is determined from the equation
\begin{equation}\label{6}
\frac{d \hat{P}}{dt}=\hat{M}\hat{P}\;.
\end{equation}
In calculations of the current-current correlation functions one
has to include the
self-correlation terms as well.~\cite{her,kor} According to this
procedure the Fourier transform of the charge-charge
$S_{NN}$ (upper sign) and spin-spin $S_{MM}$
(lower sign) correlation functions are given by
\begin{equation}\label{7}
S_{NN(MM)}(\omega)=4\sum_{N'_{\uparrow},N'_{\downarrow};N_{\uparrow},N_{\downarrow}}
\left(
N'_{\uparrow}\pm N'_{\downarrow}\right) Re \left[\frac{1}{i\omega\hat{1}-\hat{M}}\right]_{N'_{\uparrow},N'_{\downarrow};N_{\uparrow},N_{\downarrow}}
\left(N_{\uparrow}\pm N_{\uparrow}\right)p^0(N_{\uparrow},N_{\downarrow})\;,
\end{equation}
while the current-current correlation function is
\begin{equation}\label{8}
S_{II}(\omega)=S^{Sh}_{II}+S^c_{II}(\omega)\;,
\end{equation}
 where the Schottky value (the frequency independent part) is
 \begin{equation}\label{8a}
S^{Sh}_{II}=\frac{2e^2}{C^2} \sum_j \left(\frac{C_1C_2}{C_j}\right)^2\sum_{N_{\uparrow},N_{\downarrow}}
\left[\Gamma^+_{j\uparrow}(N_{\uparrow},N_{\downarrow})+\Gamma^-_{j\uparrow}(N_{\uparrow},N_{\downarrow})
+\Gamma^+_{j\downarrow}(N_{\uparrow},N_{\downarrow})+\Gamma^-_{j\downarrow}(N_{\uparrow},N_{\downarrow})\right]
p^0(N_{\uparrow},N_{\downarrow})
\end{equation}
and the second term in Eq.(13) is given by
\begin{eqnarray}\label{8b}
S^c_{II}(\omega)=\frac{4e^2}{C^2}\sum_{N'_{\uparrow},N'_{\downarrow};N_{\uparrow},N_{\downarrow}}
\{C_2\left[\Gamma^+_{1\uparrow}(N'_{\uparrow},N'_{\downarrow})-\Gamma^-_{1\uparrow}(N'_{\uparrow},N'_{\downarrow})
+\Gamma^+_{1\downarrow}(N'_{\uparrow},N'_{\downarrow})-\Gamma^-_{1\downarrow}(N'_{\uparrow},N'_{\downarrow})\right]
\nonumber\\
+C_1 \left[\Gamma^-_{2\uparrow}(N'_{\uparrow},N'_{\downarrow})-\Gamma^+_{2\uparrow}(N'_{\uparrow},N'_{\downarrow})
+\Gamma^-_{2\downarrow}(N'_{\uparrow},N'_{\downarrow})-\Gamma^+_{2\downarrow}(N'_{\uparrow},N'_{\downarrow})\right] \}\nonumber\\
\times Re \left[\frac{1}{i\omega\hat{1}-\hat{M}}\right]_{N'_{\uparrow},N'_{\downarrow};N_{\uparrow},N_{\downarrow}}\times \nonumber\\
\{\left[C_2\Gamma^+_{1\uparrow}(N_{\uparrow}-1,N_{\downarrow})-
C_1\Gamma^+_{2\uparrow}(N_{\uparrow}-1,N_{\downarrow})\right]p^0(N_{\uparrow}-1,N_{\downarrow})\nonumber\\
+\left[C_2\Gamma^+_{1\downarrow}(N_{\uparrow},N_{\downarrow}-1)-
C_1\Gamma^+_{2\downarrow}(N_{\uparrow},N_{\downarrow}-1)\right]p^0(N_{\uparrow},N_{\downarrow}-1)\nonumber\\
-\left[C_2\Gamma^-_{1\uparrow}(N_{\uparrow}+1,N_{\downarrow})-
C_1\Gamma^-_{2\uparrow}(N_{\uparrow}+1,N_{\downarrow})\right]p^0(N_{\uparrow}+1,N_{\downarrow})\nonumber\\
-\left[C_2\Gamma^-_{1\downarrow}(N_{\uparrow},N_{\downarrow}+1)-
C_1\Gamma^-_{2\downarrow}(N_{\uparrow},N_{\downarrow}+1)\right]p^0(N_{\uparrow},N_{\downarrow}+1)\}\;.
\end{eqnarray}
In the present case the Green's function $\hat{G}(\omega)=[i\omega\hat{1}-\hat{M}]^{-1}$ is
defined in the
two-dimensional space of states $(N_{\uparrow},N_{\downarrow})$, in contrast
to the previous approaches for the spinless SETs~\cite{kor,her,oth}, where the
corresponding Green's function was in the one-dimensional space (i.e. it was
tridiagonal).

\section{Fluctuations and noise in F-N-F junctions}

Let us begin our discussion with a nonmagnetic grain connected to two
ferromagnetic leads (F-N-F junction).
In Fig.1a the current-voltage characteristics are
presented for the antiparallel
and the parallel configurations.
We assumed there
the charging energy $E_c$ equal to 10.1meV
and the level spacing $\Delta E$ equal to 3meV. This means
that for the electronic bandwidth $W\approx 10eV$
the metallic grain contains $N_a\sim 10^3$-$10^4$
atoms ($\Delta E\approx W/N_a$). The temperature $T=2.3$K
is small enough to see the effects due to discrete
charging and discrete structure of the energy
levels ($k_BT\ll E_c, \Delta E$). In Fig.1b the charge and spin
accumulated on the grain are shown as a function of the bias voltage.
The steps in the charge accumulation are related to the steps of the current
as they both are caused by discrete charging effects.
As the charge accumulation is similar in both configurations,
the spin accumulation only occurs for the antiparallel configuration
and is absent in the parallel one. This is because we assumed
that the ferromagnetic
electrodes are made of the same  materials, so
$R_{1\uparrow}/R_{1\downarrow}= R_{2\uparrow}/R_{2\downarrow}$ in the
parallel configuration. The oscillations
of the spin accumulation are due to periodic blockade of the
channels for electrons with the spin $\sigma = \uparrow$
and $\sigma =\downarrow$. With
increasing $V$ the mutual balance between the currents through both junctions
leads to an increase in the spin accumulation and this gives rise to
shifts of the chemical
potentials for electrons with opposite spins. When $V$ increases further,
a new channel for spin $\sigma=\downarrow$ opens (in the present case at $V\approx 28$mV),
which leads to a decrease of the
spin accumulation as the charge accumulation is almost constant.
At $V\approx 40$mV  a new channel for electrons with $\sigma=\uparrow$
opens as well and now both the spin and charge accumulations increase and
the next cycle begins. Small steps in the curves shown in Fig.1 (and also in the
following numerical results) is due to discrete structure of the grain
energy levels.

\subsection{Charge and spin fluctuations}

 In order to study the frequency dependence of the correlation functions
$S_{NN}(\omega)$,  $S_{MM}(\omega)$ and $S_{II}(\omega)$ we perform their
spectral decomposition. Accordingly, we express the correlation functions
as the sum
\begin{equation}\label{10}
S_{XX}(\omega)=4\sum_{\lambda <0}\frac{-\lambda}{\lambda^2+\omega^2}S_{XX}(\lambda)
\end{equation}
of the components $S_{XX}(\lambda)$ in the representation of the eigenvalues $\lambda$ of
the matrix $\hat{M}$.  All eigenvalues $\lambda$
are negative, except the eigenvalue $\lambda=0$ which
corresponds to the stationary solution.
The bias dependence of the eigenvalues $\lambda$
is presented in Fig.2 for the spinless and ferromagnetic
SETs (the upper and lower parts, respectively). In the latter case
the structure  is much more complex
than in the former one because the spin degeneracy is removed in
ferromagnetic SETs.
There is also hybridization of the eigenvalues, especially strong in the
low frequency region and
for voltages close to the steps in the $I$-$V$ curves.

Calculations  of
the noise were performed numerically
for the space of states $(N_{\uparrow},N_{\downarrow})$ dynamically increasing
with $V$, and the convergence of the results was checked for the large space
of states, which in an
extreme situation included 247 states. The results of the spectral decomposition
were compared with those obtained by direct calculation of the Green's function
$\hat{G}(\omega)$
for some small values of $\omega$ (by inversion of the matrix $[i\omega\hat{1}-\hat{M}]$).~\cite{kor}

An example of the spectral decomposition of the charge-charge,
$S_{NN}(\omega)$, and spin-spin, $S_{MM}(\omega)$, correlation functions
is shown in Fig.3.
The amplitude of the spin noise $S_{MM}(\lambda)$ is
two orders of magnitude larger than the amplitude of
the charge noise $S_{N
N}(\lambda)$. From Fig.3 follows that only a few components are significant;
the others are exponentially small and can be omitted.
It would be rather difficult to extract experimentally the
contributions corresponding to  $\lambda$'s, which are close to each other.
In such a case one can approximate $S_{NN}(\omega)$ with a
single Lorentzian curve
by introducing an effective relaxation time,
\begin{equation}\label{9}
S_{NN}(\omega)=4  \frac{\tau_{charge}}{1+\omega^2\tau_{charge}^2} \mbox{var}(N)\;,
\end{equation}
as it would be for poissonian processes of independent tunneling events.
Here, var$(N)=<N^2>-<N>^2$ is the variance of $N$.
In Fig.4  we present the characteristics for the charge noise:
the variance  var$(N)$ and the inverse of the
effective relaxation time $1/\tau_{charge}$.
The voltage dependence of $1/\tau_{charge}$
is a periodic function, which has minima
at the steps of the charge accumulation and
maxima at the points corresponding to the
minima of var$(N)$. The value of $1/\tau_{charge}$ is an order
of magnitude larger than the average frequency of electron
transfer from the source to sink electrodes, $1/\tau_{tr}=I/e \approx 1$GHz.
This is because the current $I$ is determined by the
more resistive junction, for which the
electron tunneling frequency is close to $1/\tau_{tr}$, while $1/\tau_{charge}$
$(\sim E_c/$min$(R_{j\sigma}))$ is
determined by the frequency of electron tunneling through the
less resistive barrier.

Since we assumed the level spacing  $\Delta E< E_c$,  we can expect large spin fluctuations
(var$(M)$ larger than var$(N)$) and a long spin relaxation time ($\tau_{spin}>\tau_{charge}$).
The analysis of the
spin noise has been performed in a similar way as of the charge noise.
The corresponding results presented in
Fig.5 show that in the case considered
 $1/\tau_{spin}$ is 20 times smaller than $1/\tau_{charge}$.
Since $1/\tau_{spin}\sim \Delta E/$max$(R_{j\sigma})$,
its value can be even smaller for a larger
number of atoms  in the grain ($N_a\approx W/\Delta E$).
There are also large differences in var$(M)$ and $1/\tau_{spin}$ in
the antiparallel and parallel configurations. The spin
fluctuations are significantly smaller in the antiparallel configuration than
in the parallel one. This difference is due to large spin accumulation
in the antiparallel configuration, which shrinks  the space of available states for the
spin fluctuations. The small steps in var$(M)$ and $1/\tau_{spin}$,
clearly seen in Fig.5, particularly in the
antiparallel
configuration, result from the step-like changes of the spin accumulation on the
grain. Using the spectral decomposition procedure we found
that the dominant contribution to the spin noise comes
from the largest eigenvalue $\lambda_1$. This can be concluded from Fig.5b,
where the exact value of $1/\tau_{spin}$ (thick solid and dashed  curves)
is compared with the contribution
corresponding to  $\lambda_1$ (thin solid and dashed curves).

\subsection{Current noise}

The current shot noise has been calculated from the formulas
(\ref{8})-(\ref{8b}). Figure 6a shows the bias dependence of the
zero-frequency current noise
$S_{II}(\omega=0)$. The current noise is always smaller in the antiparallel
configuration
than in the parallel one. This is because in the presence of spin accumulation  the 
amplitude of fluctuations is smaller for
 the antiparallel configuration.
In Fig.6b $S_{II}(\omega=0)$ is split into two parts; a frequency independent
component
$S_{II}^{Sh}$ and a contribution  $S_{II}^c(\omega=0)$
arising from the frequency dependent part of the current noise (see Eq.(13)).
The component $S_{II}^{Sh}$ is almost constant $\approx 2eI(C_1^2+C_2^2)/C^2$
at the plateaux of the $I-V$ curve and increases with opening of new channels.
Dynamical correlations between the currents are described by $S_{II}^c(\omega)$.
Its value in the limit $\omega \to 0$ can  be positive between the  $I-V$ steps
and negative when new channels become open. This is evident 
for the antiparallel
configuration at $V\approx 26$mV when
opening a tunneling channel for electrons with $\sigma=\downarrow$ leads to 
negative dynamical correlations. This effect is almost compensated by an 
increase in $S_{II}^{Sh}$ and therefore, one gets only a small reduction 
of the current noise $S_{II}(\omega=0)$.

A large peak of $S_{II}^{Sh}$ seen in Fig.6b at the threshold
of the Coulomb blockade ($V\approx 14$mV) occurs in any asymmetrical SET 
device. It is caused by a rapid increase of a stream of tunneling electrons onto the
grain, which appears when the conducting channels open. Since the outgoing
channel has larger resistance the charge accumulation increases and part of electrons
is pushed back to the source electrode. Frequent tunneling of electrons to and
from the grain leads to an enhancement of the charge noise as well as of the
current noise (see also Fig.4).
The reduction of $S_{II}(\omega=0)$ at $V\approx 14$mV and $V\approx 40$mV
is the effect of the opening of new current channels at the steps of the $I-V$
curves.

Frequency dependence of the $S_{II}^c(\omega)$  component of the
current noise is presented in Fig.7a for different voltages. One can clearly
distinguish between two different relaxation processes
in the current noise; respectively in the low and high frequency ranges.
An interesting case is that  for $V =36$ mV (and also $V =42$ mV), where
anticorrelations between currents occur in a low frequency
limit (with $S_{II}^c<0$) and correlations for higher
frequencies ($S_{II}^c>0$).

In order to determine the effective relaxation time $\tau_{II}$ we have performed
a detail analysis of $S_{II}^c(\omega)$ as a function of $\omega$.
The shape of $S_{II}^c(\omega)$ suggests that there are only a few
relaxation processes with well separated relaxation times $\tau_{II}$. Therefore,
one can assume that $S_{II}^c(\omega)$ is a superposition of a few Lorentzian curves.
The relaxation time for each process is
given by $1/\tau_{II}^2=3\omega_0^2$, where $\omega_0$ corresponds to the
maximal slope of $S_{II}^c$ and can be determined from $d^2S_{II}^c/d\omega^2=0$.
The results  are presented in Fig.7b by open circles.
One can distinguish two distinct relaxation times, one in the high and another one in the
low frequency ranges. In a wide voltage range
the corresponding relaxation times are very close
to the effective relaxation times for the charge and spin noise (compare
with the solid curves in Fig.7b). Some deviations occur for voltages where  new
tunneling channels become open. In a certain voltage range there are three relaxation times
(e.g. at $V\approx 36$mV - see also the curve with squares in Fig.7a). This analysis
shows that both charge and spin fluctuations are relevant for the shot noise of
the current in ferromagnetic SETs.

\subsection{Paramagnetic limit}

An interesting limit of the F-N-F junction discussed above is the case
where  all electrodes are nonmagnetic. It is important to note that large
spin fluctuations are present in such the case as well.
Fig.8 shows characteristic frequency dependence of
$S_{II}^c$ as well as of $S_{NN}$ and $S_{MM}$. Although the
amplitude of the spin noise is very large
(in this case it is four orders of magnitude larger than the charge noise),
it is not seen in the current noise.
In a nonmagnetic system the
tunnel resistances for both spin directions are the same. This
implies that the components of the current-current correlation
functions obey the relation $S^c_{I_{j\uparrow}I_{j\uparrow}}(\lambda) = S^c_{I_{j\downarrow}I_{j\downarrow}}(\lambda)$.
For those eigenvalues $\lambda$ which correspond to the spin
noise we have also 
$S^c_{I_{j\uparrow}I_{j\uparrow}}(\lambda) = -S^c_{I_{j\uparrow}I_{j\downarrow}}(\lambda) < 0$.
This means that the tunneling events with the frequency
$1/\tau_{spin}$ are anticorrelated  for electrons with the same spin
and correlated  for electrons with the opposite spin. Therefore,   the spin
component of the total current noise  is completely compensated for each
junction in a nonmagnetic SET.

The asymmetry between the tunneling channels for electrons
with the opposite spins leads to activation of the spin component in
the current noise. We extracted from $S^c_{II}$ the components
$S^c_{II\;charge}$ and $S^c_{II\;spin}$
corresponding to the charge and the spin noise, respectively. Their values in
the zero-frequency limit are determined from
$S^c_{II\;charge}(\omega=0)=-4\sum_{\lambda <\lambda_0}S_{II}(\lambda)/\lambda$ and
$S^c_{II\;spin}(\omega=0)=-4\sum_{\lambda_0 <\lambda <0}S_{II}(\lambda)/\lambda$, where
$\lambda_0= -1/\sqrt{\tau_{charge}\;\tau_{spin}}$. The results are
presented in Fig.9 as a function of the parameter $p$, where 
$p_1=p_2$ with $p_j=(R_{j\uparrow}-R_{j\downarrow})/(R_{j\uparrow}+R_{j\downarrow})$ for $j=1$, 2. 
It can  be seen that the charge component is almost 
constant whereas the spin component increases with $p$ and
for $p\to 1$ can be  much larger than the charge component.

\section{Enhancement of the noise in F-N-N junctions}

Let us consider now the electron transport through a nonmagnetic metallic
grain connected to a nonmagnetic lead on one side and to a ferromagnetic lead
on the other side (F-N-N junction). When the ferromagnetic electrode is
a strong ferromagnet with only one spin subband (say for spin
$\sigma=\downarrow$) partially filled, then the
electrons with $\sigma=\uparrow$ can tunnel onto the grain from the
nonmagnetic electrode, but the outgoing $\sigma=\uparrow$ channel
from the grain to the
ferromagnetic electrode is blocked ($R_{2\uparrow}=\infty$). Such a double
junction can be considered as a system
of coupled SET (for $\sigma=\downarrow$) and a single electron box (SEB)
(for $\sigma=\uparrow$).
Nonmagnetic coupled SET-SEB devices were recently analyzed theoretically and it
was shown that
if the coupling is strong,
one can have a negative differential resistance (NDR).~\cite{ndr,delft}
Heij et al.~\cite{delft} constructed recently the SET-SEB device based
on two electrostatically coupled metallic grains, and found
NDR indeed.
In our case  we have only a single grain, but two coupled electron
channels.
In the $I$-$V$ curve shown in Fig.10 we see NDR with a sequential drop of the current.
With increasing $V$ the number of electrons with $\sigma=\uparrow$ on
the grain increases (see the spin accumulation in Fig.10b), which results in an increase
of the local chemical potential for $\sigma=\uparrow$ electrons and a drop of
the current. The maximum drop of $I$ occurs at the maximum of the spin
accumulation (at $V\approx 26$mV). Opening of the next charge channel ($N\to N+1$) for conducting
electrons ($\sigma=\downarrow$) reduces the spin accumulation and increases the
current. Small steps in the charge and spin accumulations are the effect of
opening of spin channels, i.e. for 
$(N_{\uparrow},N_{\downarrow})\to  (N_{\uparrow}\pm 1,N_{\downarrow}\mp 1)$.

The noise in magnetic SET-SEB devices has been analyzed with the same method
as before and the results for the charge
and spin noise are presented in Fig.11. The magnetic fluctuations
increase when a new channel opens and then they drop again with
increasing voltage. The peaks in
var$(M)$ correspond to the maxima of the effective relaxation time $\tau_{spin}$ (the
minima of $1/\tau_{spin}$ shown in Fig.11b). The spin noise
has a significant influence on the current
noise shown in Fig.12. The large increase of the current noise (much over the value $S_{Poisson}$)
takes place for the range of NDR, which occurs for voltages
corresponding to large spin fluctuations.
We performed the spectral decomposition of the current noise $S_{II}^c(\omega)$.
The voltage dependence of the eigenvalues $\lambda$ did not show any peculiarities in the
range of NDR. The spectral weight is, however, shifted to lower frequencies.
The frequency dependence of $S_{II}^c(\omega)$, shown in Fig.13,
exhibits strong increase
of the low frequency contribution for the voltages where NDR occurs. Outside the NDR range
the low frequency component of the current noise is minor.

Recently, Iannaccone et al.~\cite{iana} studied resonant tunneling
through a double barrier GaAs structure and showed an enhancement of the shot
noise in the region of NDR.  The NDR effect was  due to charge
accumulation in the central well (the effect predicted by Ricco and Azbel~\cite{azbel}).
The enhancement of the shot noise was also observed in resonant
tunneling through a double barrier
in magnetic field,~\cite{brown} although NDR was a result of changes
in electronic structure in high magnetic field (formation of Landau levels).
Iannaccone et al.~\cite{iana} analyzed the enhancement of the shot
noise  in the framework of the generation-recombination
approach~\cite{vliet,davies,iana97,ziel} in terms of the effective generation ($g$)
and the recombination ($r$) rates and their relaxation times $\tau_g$
and $\tau_r$, respectively. Such the approach
is equivalent to two channel approximation.~\cite{kor,her,oth,eto}
In the present studies we used the method which takes into account
multichannel (and multilevel) tunneling processes and
therefore we have insight into the microscopic processes related to the
tunneling current and its noise.

\section{Final remarks}

In summary, we have performed theoretical analysis of the frequency dependent
current shot noise in ferromagnetic
tunnel junctions in the limit of sequential tunneling. We
showed that apart from charge fluctuations there are also strong spin fluctuations,
which influence the current noise. The spin noise is significant at
low frequencies,
while the charge noise at higher frequencies.
We predict that  two
distinct relaxation times $\tau_{spin}$ and $\tau_{charge}$ should be seen in
frequency measurements of the current noise in magnetic
tunnel junctions.

We have also studied the magnetic SET-SEB device. The spin accumulation
effect can lead in this case to a negative differential resistance,
which occurs only when the spin accumulation
increases, i.e., when the chemical potential of the {\it conducting}
electrons increases due to accumulation. Spin fluctuations are then activated, which leads to a
strong enhancement of the low frequency component of the current noise.

The numerical results have been calculated at a low temperature
($k_BT\ll E_c, \Delta E$), where all features due to discreteness of the system are
clearly seen. We have also performed calculations for higher temperatures. The small 
steps, corresponding
to the opening of the magnetic channels, are washed out when $T\to \Delta E/k_B$. 
However, the charge and spin
fluctuations, var$(N)$ and var$(M)$, and the corresponding
relaxation times, $\tau_{charge}$ and
$\tau_{spin}$, remain  almost unchanged. In F-N-F junctions (as in Fig.1) the tunnel
magnetoresistance, defined as $TMR=I_P/I_{AP}-1$ (where $I_P$ and $I_{AP}$ are the currents in
the parallel and in the antiparallel configurations), decreases with increasing
temperature (see also Ref.[\onlinecite{mar}]). For  $T> \Delta E/k_B$,
the physical quantities become less dependent on the voltage.  The values of var$(N)$ and
var$(M)$ increase with $T$, as expected. The charge and spin fluctuations
are still well separated in the different frequency ranges. At $T=34.8$K
the zero-frequency current noise
$S_{II}(\omega=0)\approx 0.8S_{Poisson}$. The reduction is due to Coulomb
correlations, which are still large as $T<E_c/k_B=117$K. The component $S_{II}^{Sh}$ is,
however, much larger than $S_{Poisson}$, whereas
the frequency dependent part $S_{II}^c(\omega)$ is negative with a large
amplitude. Its low and high frequency components, corresponding to the spin
and the charge fluctuations, are also well defined.

\acknowledgments
	The paper is supported from the State Committee for Scientific Research
Republic of Poland within Grant No.~2 P03B 075 14.

\newpage
\vskip 14cm
\epsfxsize=14cm
\epsffile{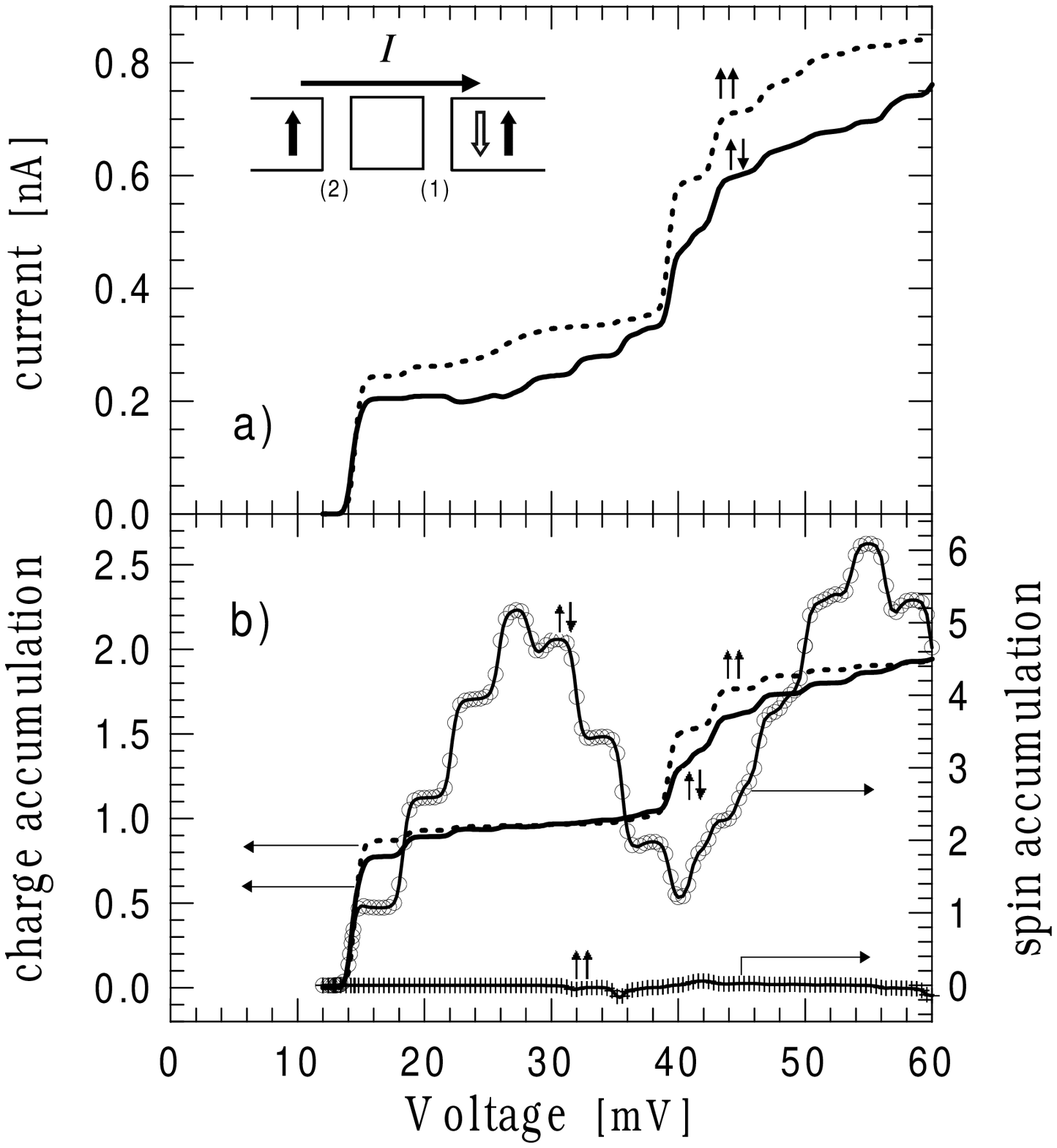 }
\hskip 1cm
\begin{figure}
\caption{Voltage dependences of the current (a) and
charge (solid and dashed curves)
and spin (curves with circles and crosses) accumulation (b)
in the antiparallel and the parallel configurations.
The parameters assumed here are:
$R_{2\uparrow}=240$M$\Omega$, $R_{2\downarrow}=60$M$\Omega$,
$R_{1\uparrow}=2$M$\Omega$ and $R_{1\downarrow}=8$M$\Omega$ for
the antiparallel configuration
($R_{2\uparrow}=240$M$\Omega$, $R_{2\downarrow}=60$M$\Omega$,
$R_{1\uparrow}=8$M$\Omega$ and $R_{1\downarrow}=2$M$\Omega$
for the parallel configuration),
$C_2=6.6$aF, $C_1=1.32$aF, $\Delta E=3$meV and  $T=2.3$K.
The inset in part (a) shows the scheme of the system 
(electrons flow from right to left).}
\label{f1}
\end{figure}

\newpage
\vskip 14cm
\epsfxsize=14cm
\epsffile{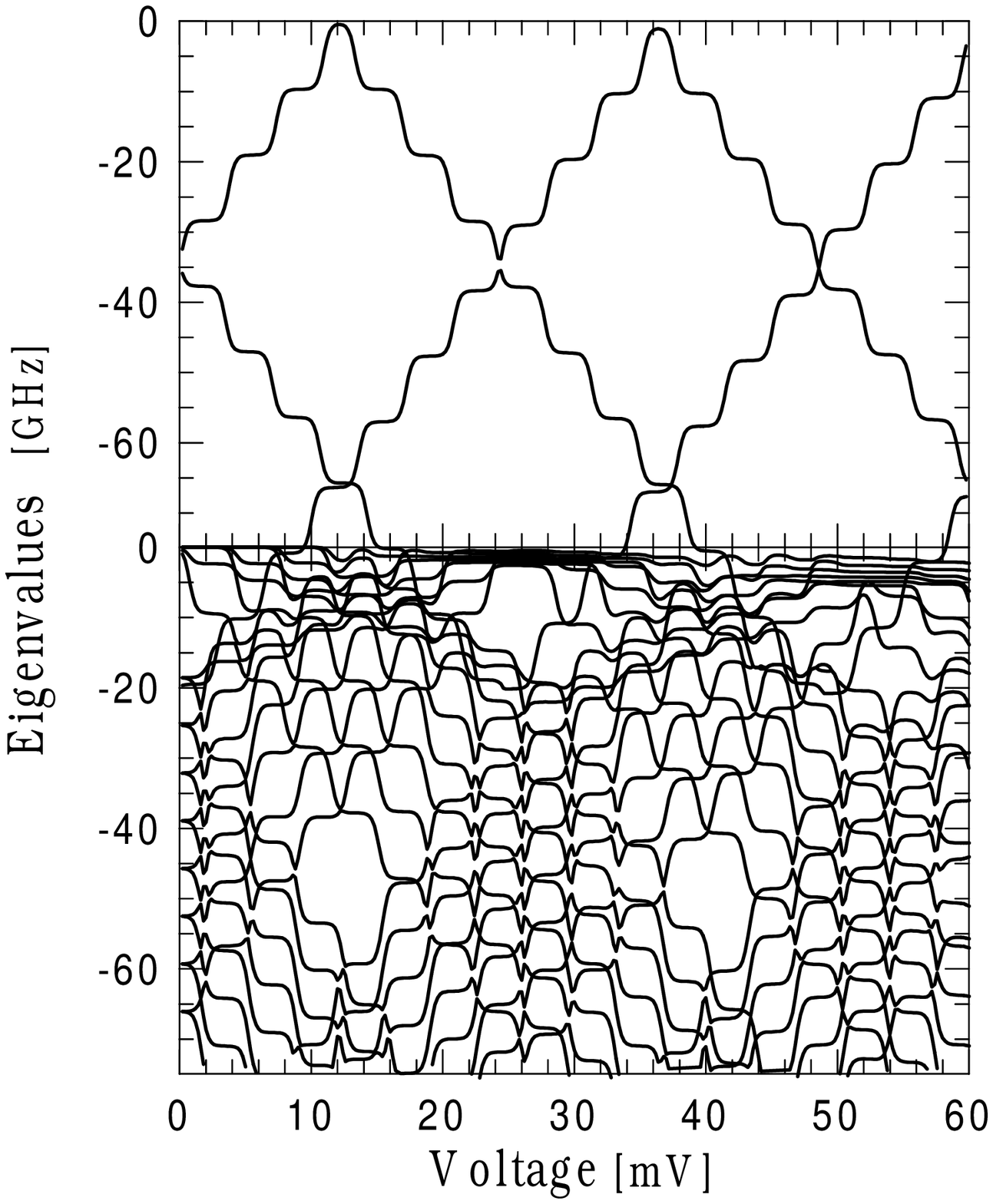 }
\hskip 1cm
\begin{figure}
\caption{Voltage dependence of the eigenvalues $\lambda$ of the matrix $\hat{M}$
for the spinless  (the upper part) and ferromagnetic (bottom part) SETs.
The parameters assumed in (a) are:
$R_{2}=240$M$\Omega$, $R_{1}=2$M$\Omega$, $C_2=6.6$aF,
$C_1=1.32$aF and $\Delta E= 3$meV, while in (b) are the same as
in Fig.1 for the antiparallel configuration.
The space of states is confined to $-5\le N\le 5$ for the spinless SET and $-1\le N\le 4$,
 $-8\le M \le 8$ for the ferromagnetic SET.}
\label{f2}
\end{figure}

\newpage
\vskip 14cm
\epsfxsize=14cm
\epsffile{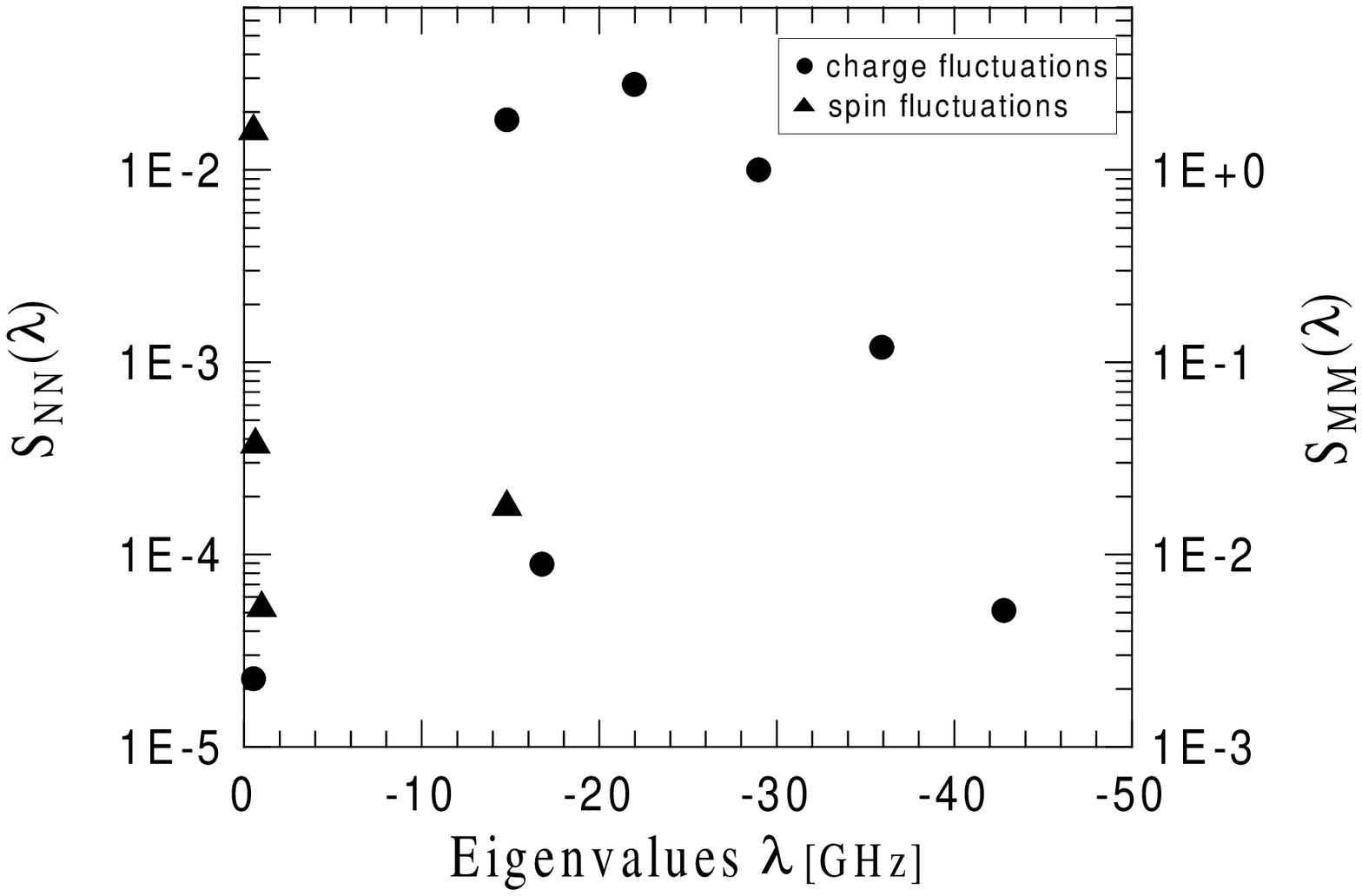 }
\hskip 1cm
\begin{figure}
\caption{The components of the spectral decomposition
of the charge-charge and the spin-spin  correlation
functions: $S_{NN}(\lambda)$ (dots) and $S_{MM}(\lambda)$ (triangles),
determined for $V=25$mV. The other parameters are as in Fig.1
for the antiparallel configuration.}
\label{f3}
\end{figure}

\newpage
\vskip 14cm
\epsfxsize=14cm
\epsffile{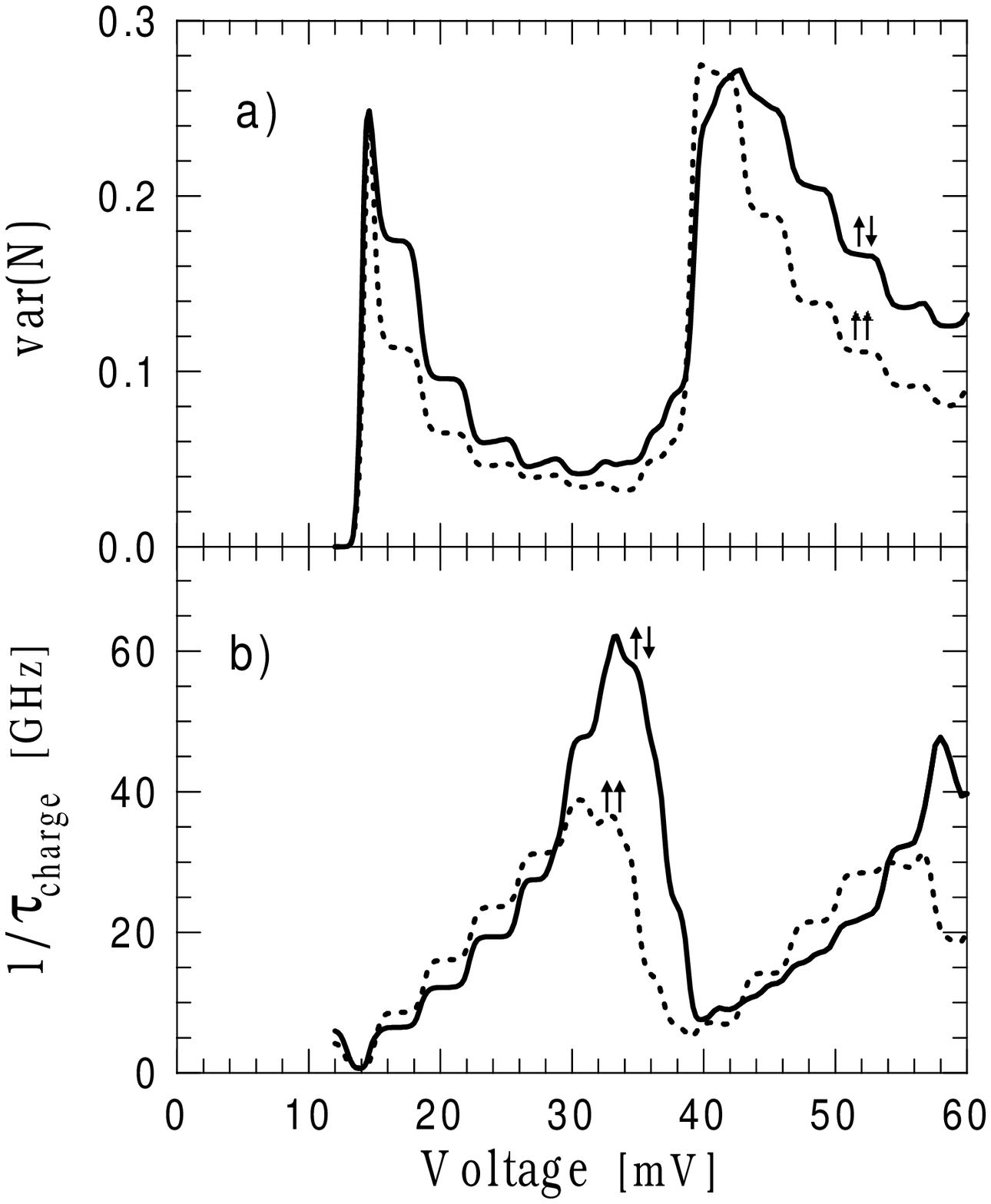 }
\hskip 1cm
\begin{figure}
\caption{Voltage dependence of the variance of charge fluctuations,
var$(N)$, (a) and of
the inverse effective relaxation time, $1/\tau_{charge}$,
(b) for the parallel and antiparallel configurations. The
parameters are the same as in Fig.1.}
\label{f4}
\end{figure}

\newpage
\vskip 14cm
\epsfxsize=14cm
\epsffile{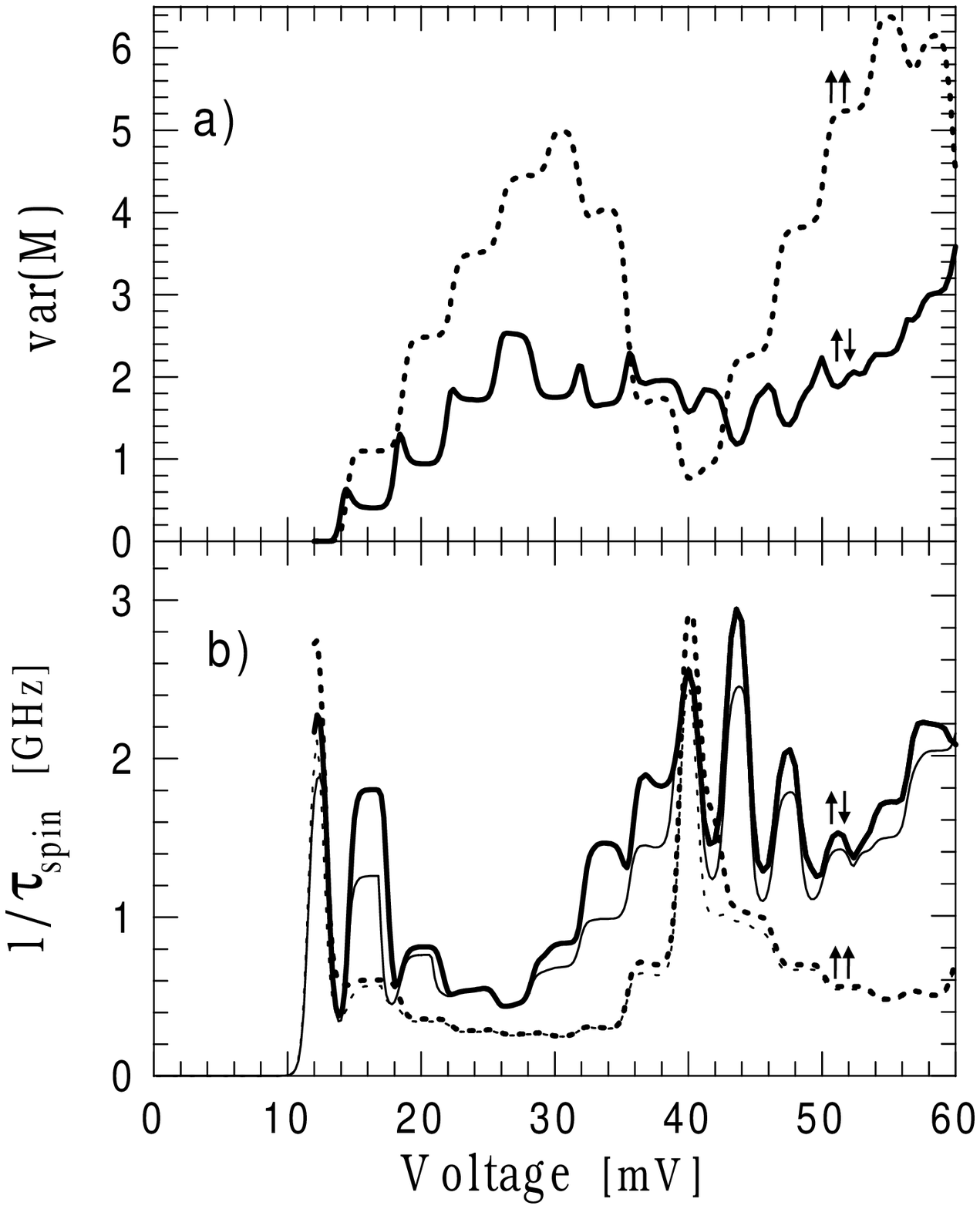 }
\hskip 1cm
\begin{figure}
\caption{Voltage dependence of the variance var$(M)$ of the spin fluctuations
(a) and of the
inverse effective relaxation time $1/\tau_{spin}$
(b) in the antiparallel and the parallel configurations.
The parameters are the same as in Fig.1. The thin (solid and dashed)
curves in part (b) present contribution
corresponding to $\lambda_1$ , where $\lambda_1$ is the largest eigenvalue
of the matrix $\hat{M}$.}
\label{f5}
\end{figure}

\newpage
\vskip 14cm
\epsfxsize=14cm
\epsffile{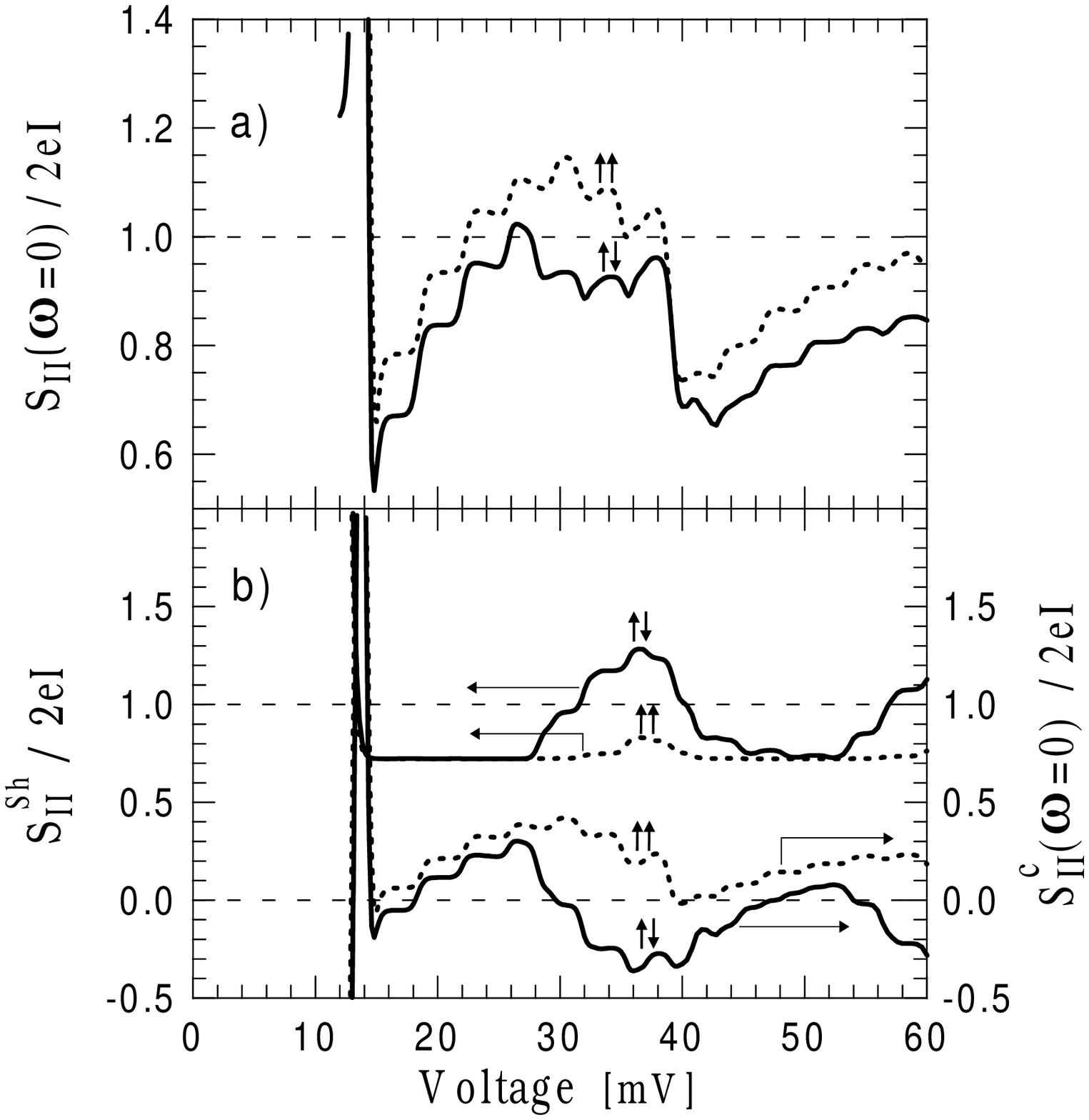 }
\hskip 1cm
\begin{figure}
\caption{Voltage dependence of the current shot noise at $\omega=0$ (a)
for the system defined in  Fig.1.
In part (b)  $S_{II}(\omega=0)$
is split into two components:  $S_{II}^{Sh}$ (upper curves)
and $S_{II}^c(\omega=0)$
(lower curves).}
\label{f6}
\end{figure}

\newpage
\vskip 14cm
\epsfxsize=14cm
\epsffile{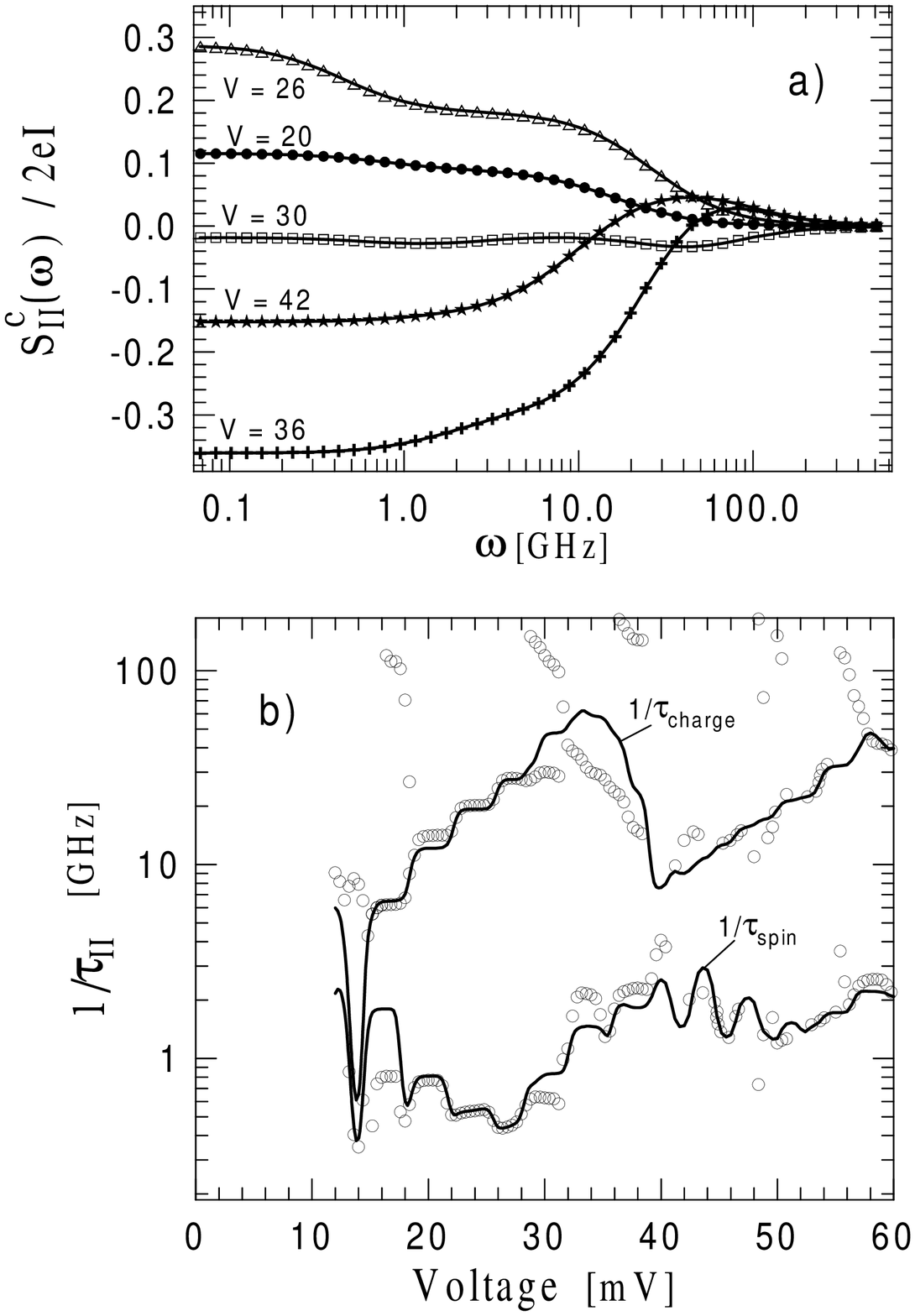 }
\hskip 1cm
\begin{figure}
\caption{(a) Frequency dependence of the current shot noise $S_{II}^c(\omega)$
in the system defined in Fig.1 for the antiparallel configuration.
(b) Inverse of the relaxation time for the current noise (open circles) as a
function of $V$. For comparison the inverse of the effective relaxation times
$1/\tau_{charge}$
and $1/\tau_{spin}$ for the
charge and the spin noise are shown by the solid upper
and the lower curve, respectively (the same curves as the solid ones in Fig.4b and Fig.5b).}
\label{f7}
\end{figure}

\newpage
\vskip 14cm
\epsfxsize=14cm
\epsffile{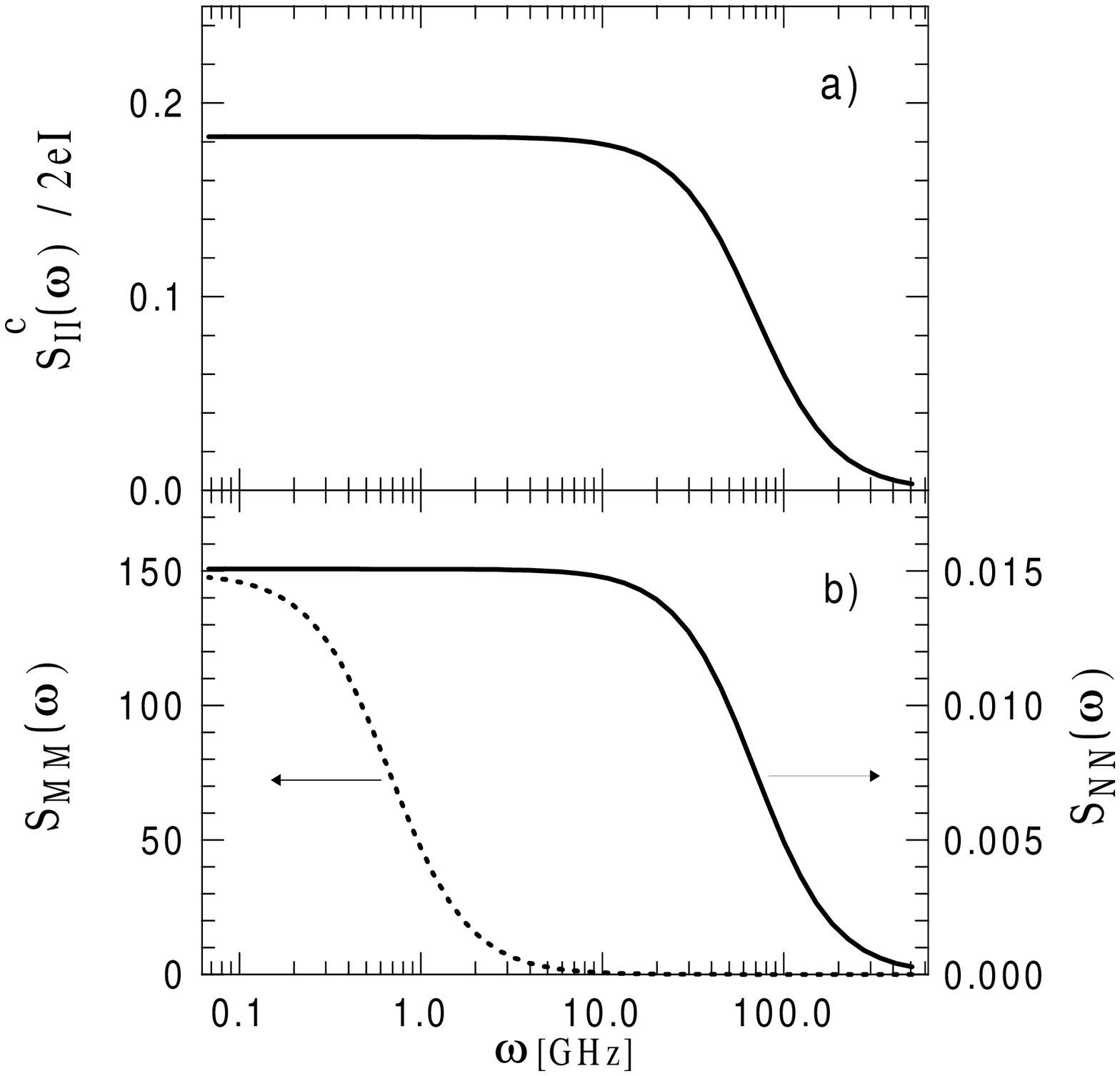 }
\hskip 1cm
\begin{figure}
\caption{Frequency dependence of the current noise $S_{II}^c(\omega)$ (a), and
of the spin and charge noise (b) at $V=26$mV in a nonmagnetic junction.
The parameters assumed are: $R_{1\uparrow}=R_{1\downarrow}=2$M$\Omega$, 
$R_{2\uparrow}=R_{2\downarrow}=60$M$\Omega$,
$C_2=6.6$aF, $C_1=1.32$aF, $\Delta E=3$meV and $T=2.3$K.}
\label{f8}
\end{figure}

\newpage
\vskip 14cm
\epsfxsize=14cm
\epsffile{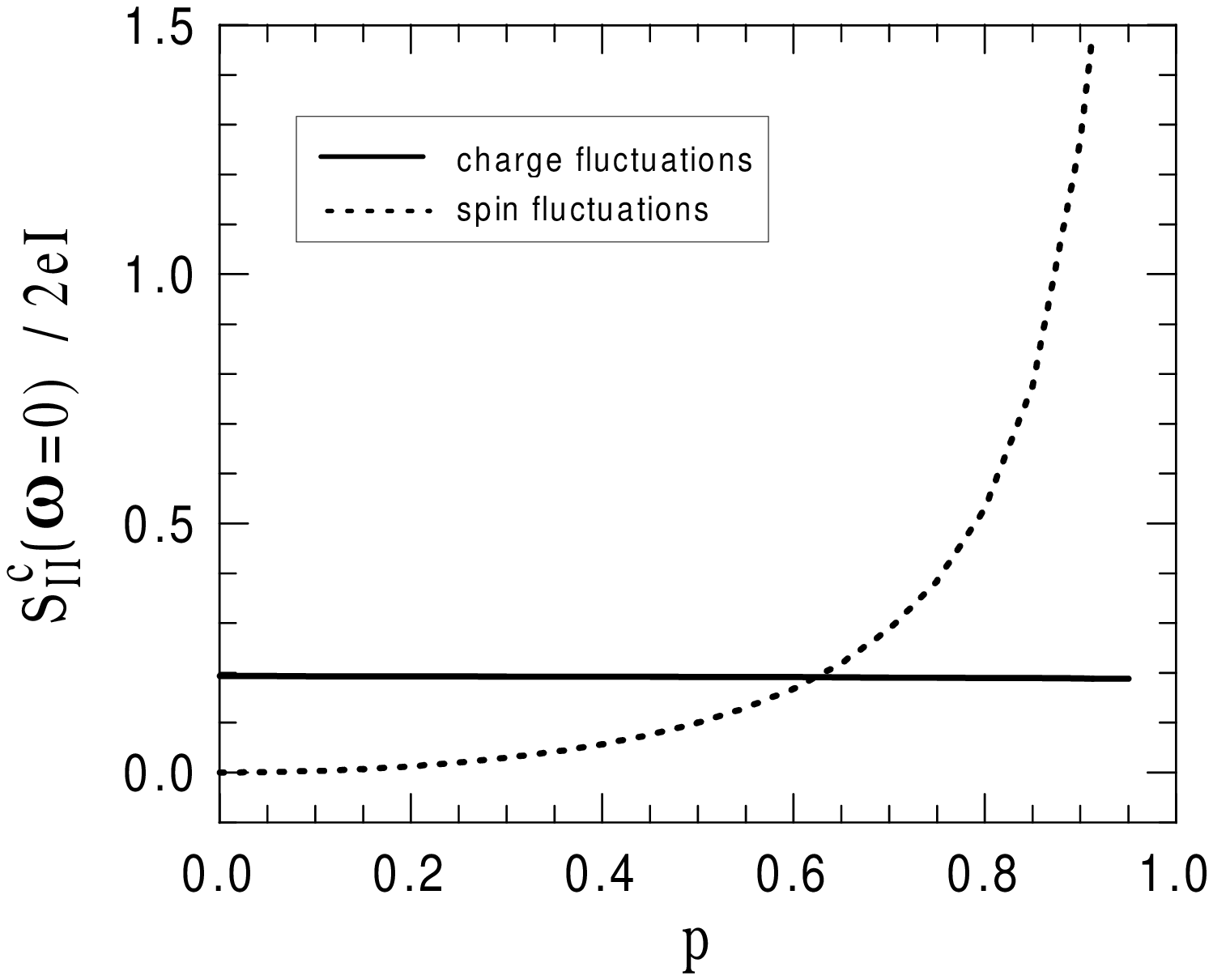 }
\hskip 1cm
\begin{figure}
\caption{Two components of $S_{II}^c(\omega=0)$ corresponding to the charge
and spin noise plotted as a function of 
$p=(R_{1\uparrow}-R_{1\downarrow})/(R_{1\uparrow}+R_{1\downarrow})= (R_{2\uparrow}-R_{2\downarrow})/(R_{2\uparrow}+R_{2\downarrow})$.
The other parameters assumed here are:
$R_{2\downarrow}=60$M$\Omega$, $R_{2\uparrow}=(1+p)/(1-p)R_{2\downarrow}$, 
$R_{1\downarrow}=2$M$\Omega$,
$R_{1\uparrow}=(1+p)/(1-p)R_{1\downarrow}$, $C_2=6.6$aF, $C_1=1.32$aF,
$\Delta E=3$meV and $T=2.3$K.}
\label{f9}
\end{figure}

\newpage
\vskip 14cm
\epsfxsize=14cm
\epsffile{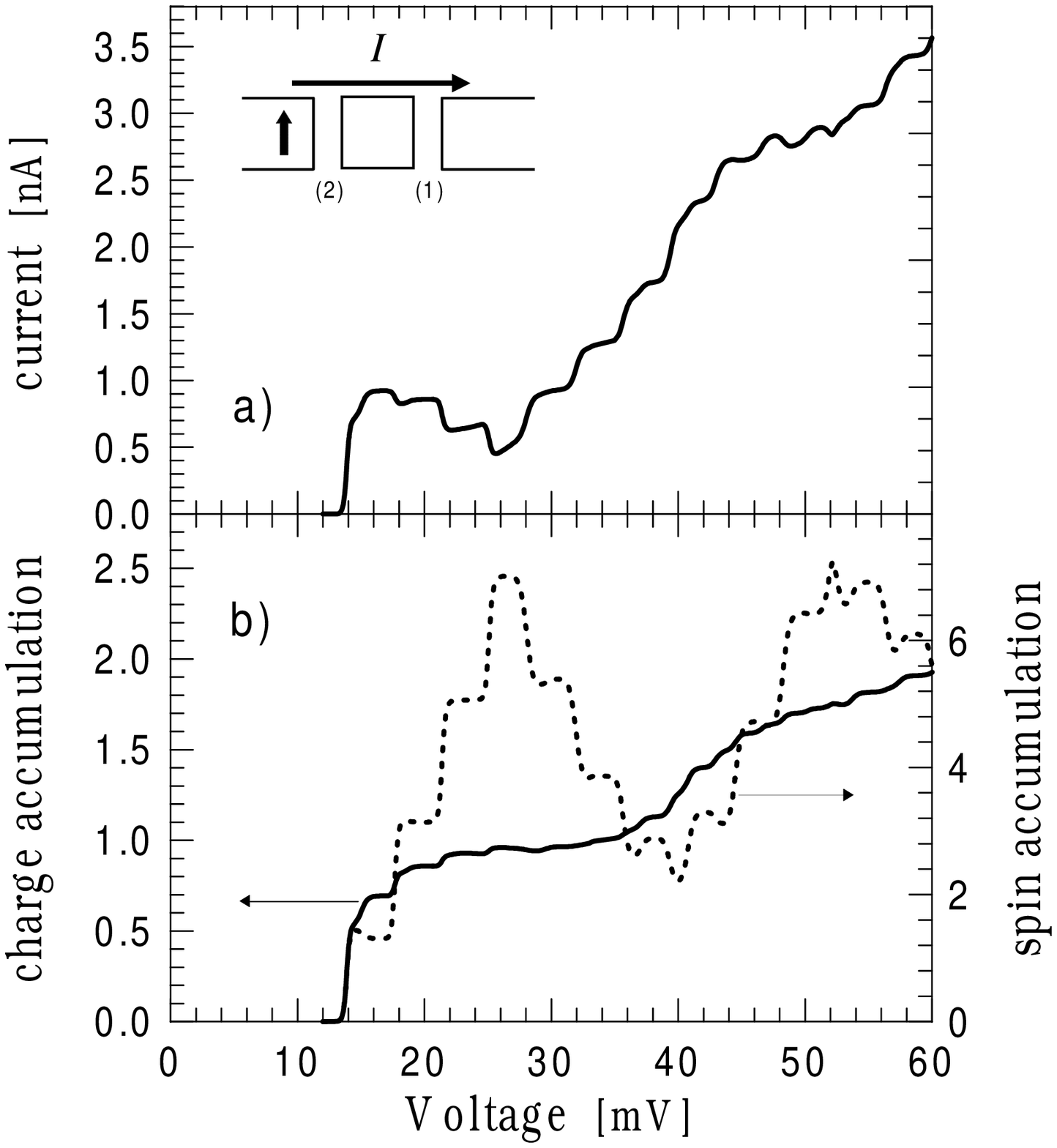 }
\hskip 1cm
\begin{figure}
\caption{The $I-V$ curve (a) and  the charge and spin
accumulation (b) for a F-N-N
junction. The parameters assumed are:
$R_{2\uparrow}=\infty$, $R_{2\downarrow}=9$M$\Omega$,
$R_{1\uparrow}=R_{1\downarrow}=2$M$\Omega$,
$C_2=6.6$aF, $C_1=1.32$aF, $\Delta E=3$meV and
$T=2.3$K. The inset shows the scheme of the system (electrons
flow from the nonmagnetic to the ferromagnetic electrode).}
\label{f10}
\end{figure}

\newpage
\vskip 14cm
\epsfxsize=14cm
\epsffile{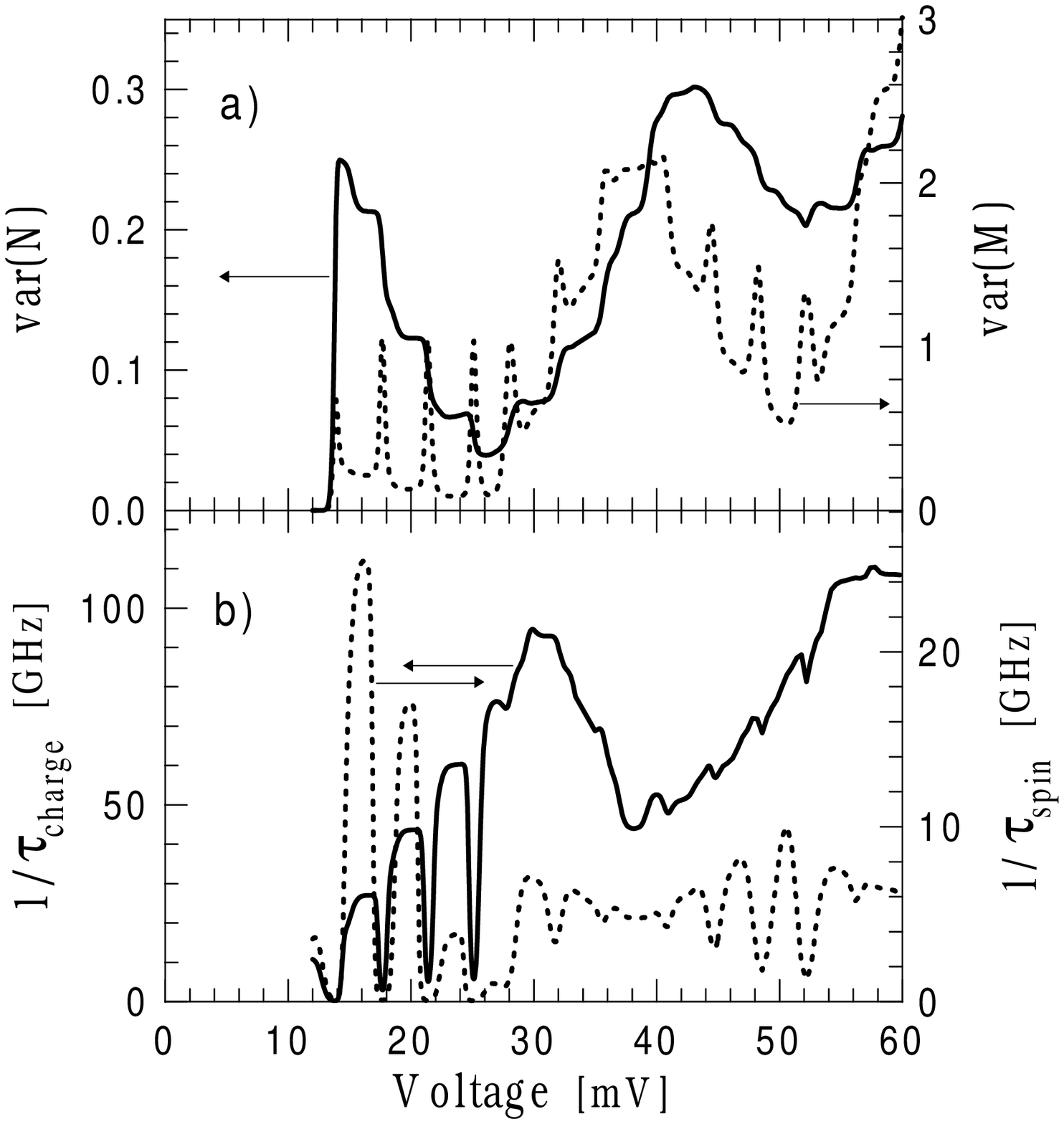 }
\hskip 1cm
\begin{figure}
\caption{The variance of the charge  and spin fluctuations (a) and
the inverse effective relaxation times  $1/\tau_{charge}$ (solid curve) and
$1/\tau_{spin}$ (dashed curve) (b) as a function of the bias voltage
for the system as in Fig.10.}
\label{f11}
\end{figure}

\newpage
\vskip 14cm
\epsfxsize=14cm
\epsffile{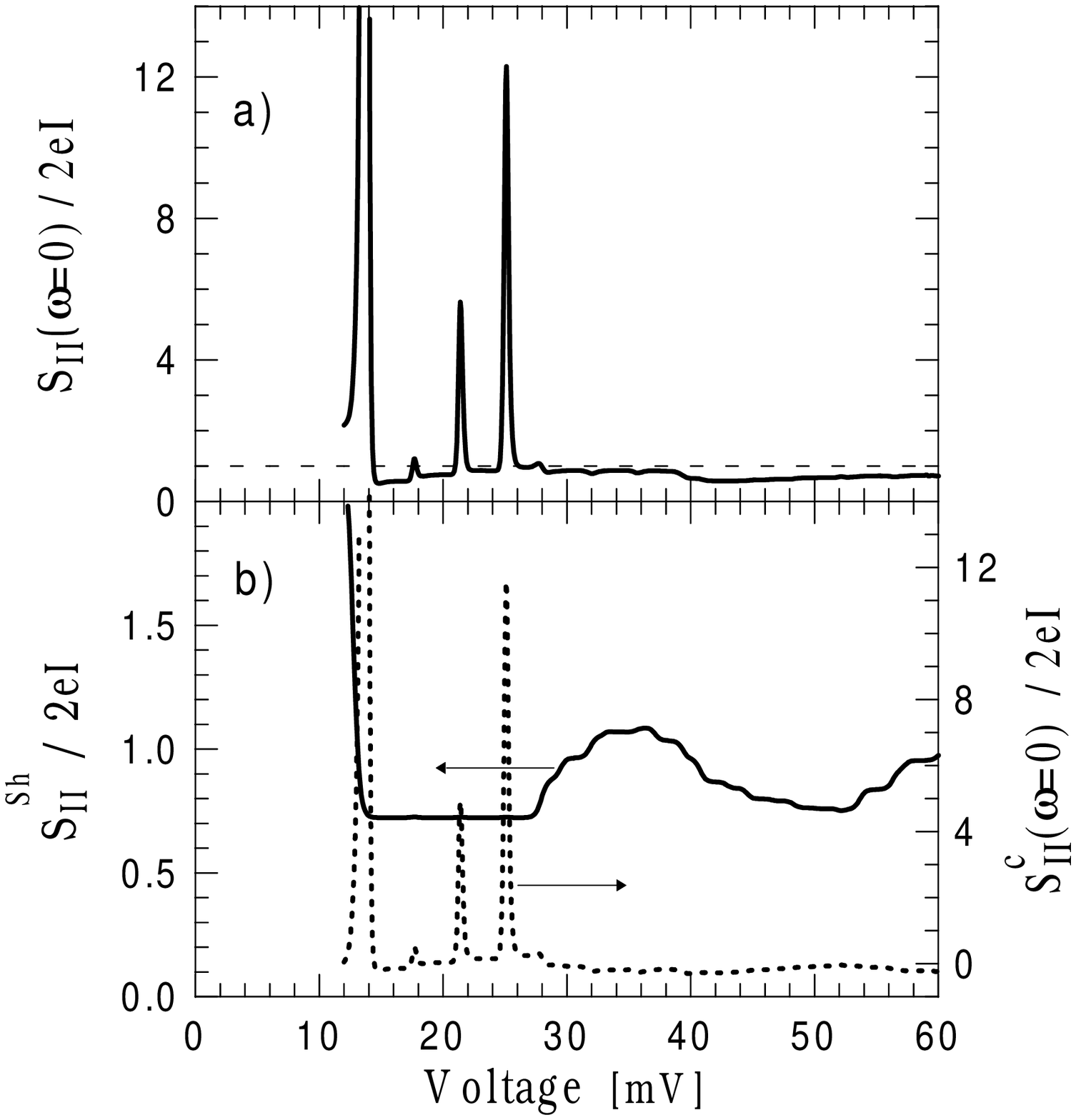 }
\hskip 1cm
\begin{figure}
\caption{Voltage dependence of the zero-frequency current shot noise (a)
in the system defined in Fig.10. In part (b)
two components of the shot noise $S_{II}^{Sh}$ (solid curve) and $S_{II}^c(\omega=0)$
(dashed curve) are presented.}
\label{f12}
\end{figure}

\newpage
\vskip 14cm
\epsfxsize=14cm
\epsffile{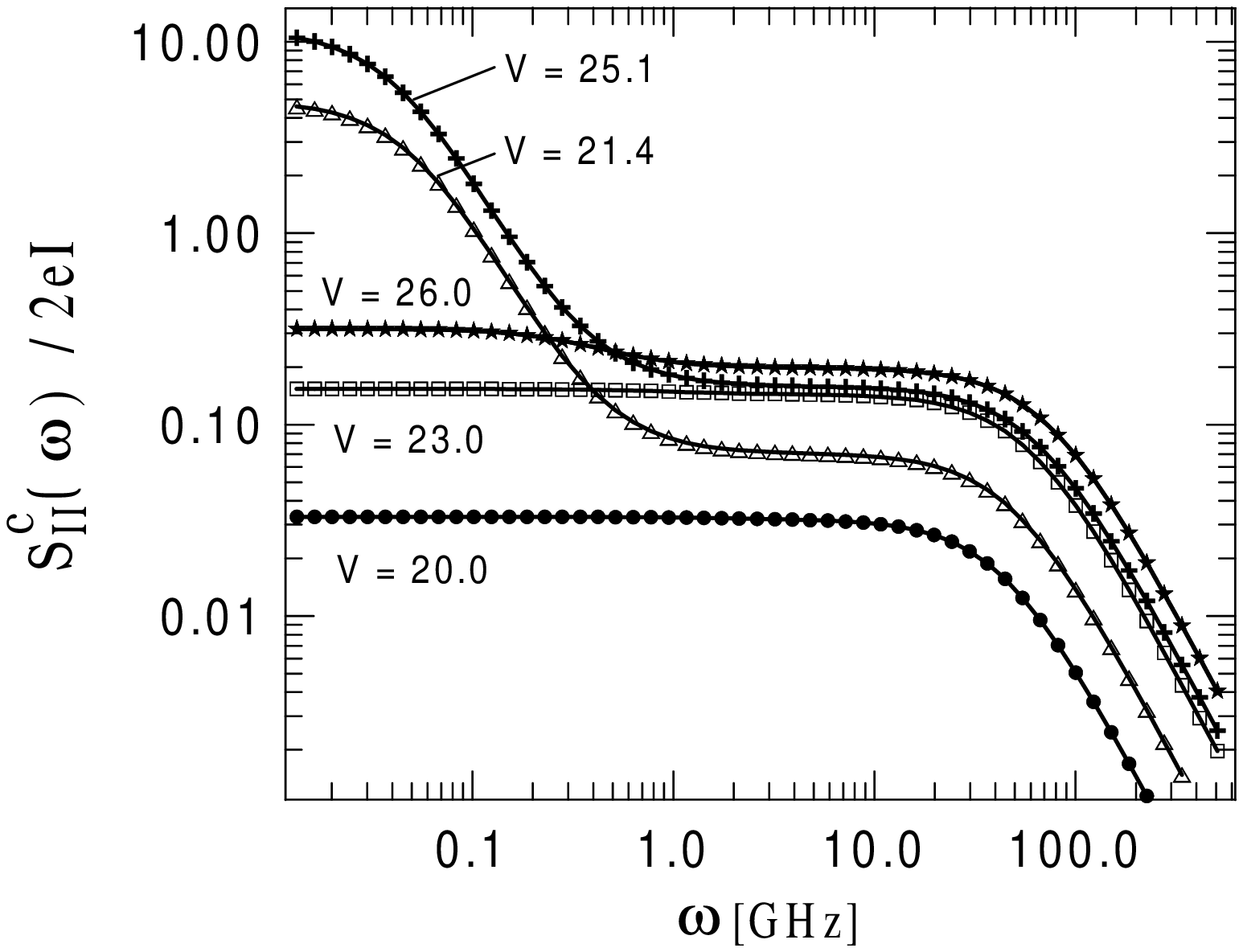 }
\hskip 1cm
\begin{figure}
\caption{Frequency dependence of the current noise $S_{II}^c(\omega)$
in the system as in Fig.10 for indicated voltages.
The curves for $V=21.4$mV and
$V=25.1$mV correspond to the maxima of the current shot
noise seen in Fig.12 in the range of NDR.}
\label{f13}
\end{figure}

\end{document}